\documentclass[aps,prd,nopacs,amsmath,amssymb,graphics,onecolumn,preprint,12pt]{revtex4}
\usepackage{graphicx}
\usepackage{dcolumn}
\usepackage{bm}

\newcommand{\be}{\begin{equation}}
\newcommand{\ee}{\end{equation}}
\newcommand{\ba}{\begin{eqnarray}}
\newcommand{\ea}{\end{eqnarray}}
\newcommand{\ban}{\begin{eqnarray*}}
\newcommand{\ean}{\end{eqnarray*}}

\begin{document}
\title{The First Detection of Gravitational Waves}
\author{Andrzej Kr\'{o}lak\footnote{Electronic address: krolak@impan.pl},
Mandar Patil\footnote{Electronic address: mpatil@impan.pl},
}

\affiliation{Institute of Mathematics of Polish Academy of Sciences \\
Sniadeckich 8, Warsaw 00-656, Poland
}


\begin{abstract}

This article deals with the first detection of gravitational waves by the 
advanced Laser Interferometer Gravitational Wave Observatory (LIGO) detectors on
14 September 2015, where the signal was generated by two stellar mass black holes with
masses 36 $ M_{\odot}$ and 29 $ M_{\odot}$ that merged to form a 62 $ M_{\odot}$ black hole,
releasing 3 $M_{\odot}$ energy in gravitational waves, almost 1.3 billion years ago.
We begin by providing a brief overview of gravitational waves, 
their sources and the gravitational wave detectors. We then describe in 
detail the first detection of gravitational waves from a binary black hole merger.
We then comment on the electromagnetic follow up of the detection event with various telescopes.
Finally, we conclude with the discussion on the tests of gravity and fundamental physics 
with the first gravitational wave detection event.

\end{abstract}

\maketitle

\section{ An Overview to Gravitational Waves and Its Sources}
\begin{figure}
\begin{center}
\includegraphics[width=12 cm]{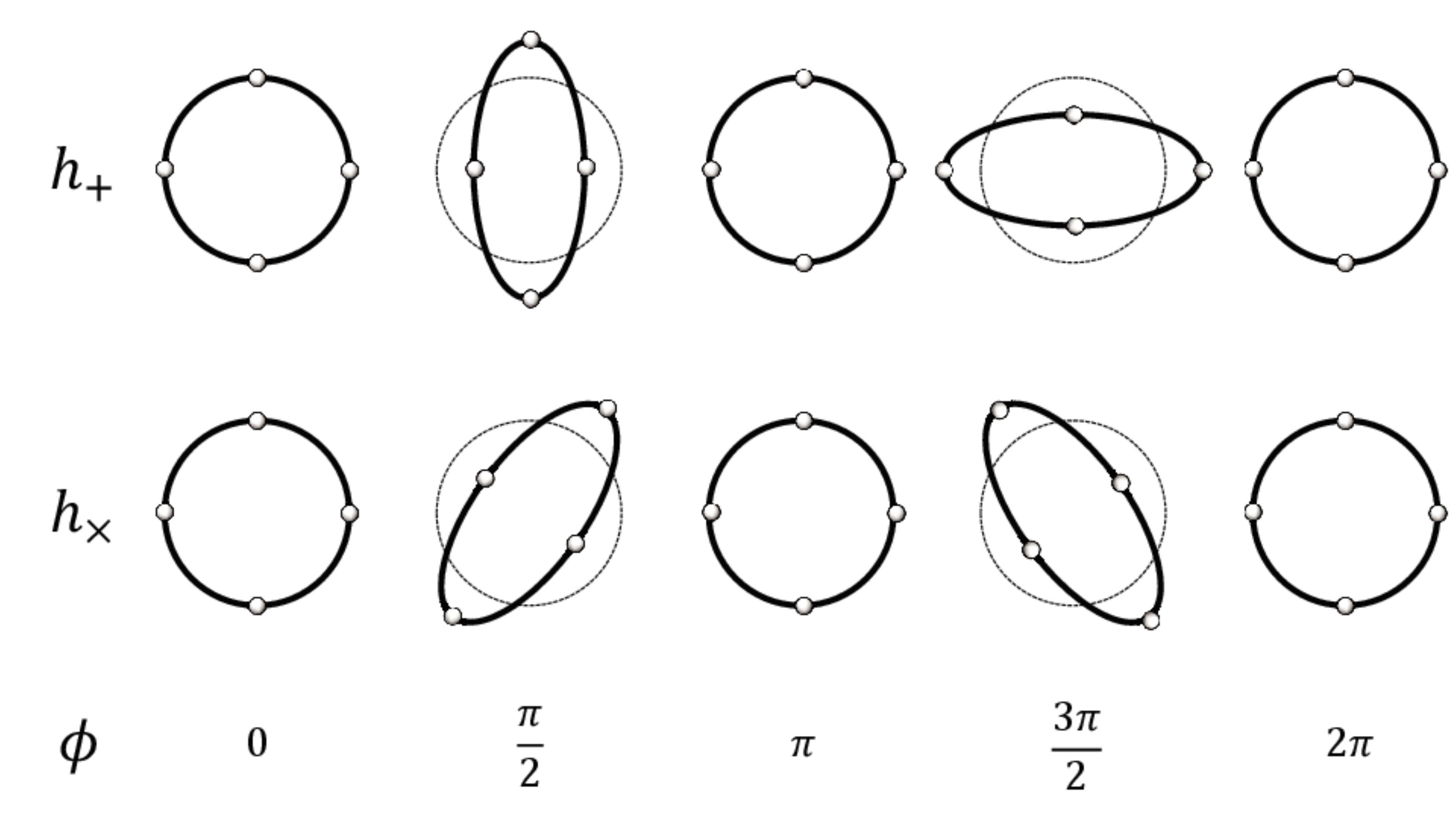}
\caption{ The effect of the passage of the linearly polarised gravitational wave through the ring of particles in the direction orthogonal to the plane is depicted in this figure. The circle shrinks in one direction and expands in the orthogonal direction during the crest and vice versa during the trough. This effect occurs along the directions
oriented 45$^{\circ}$ with respect to one another for plus and cross polarizations denoted by $h_{+}$  and $h_{\times}$, respectively. (Image credit: Eur. Phys. J. Plus 132 (2017) no. 1, 10) }\label{fig1}
\end{center}
\end{figure}

Gravitational waves were detected for the first time on 14 September  2015, which is not only the landmark discovery on its own, but also opened up a completely new observational window through which to explore the universe. It is the upshot of the tireless efforts of thousands of physicists across the globe over several decades on numerous fronts: theoretical and experimental as well as computational. As~was correctly pointed out by the Laser Interferometer Gravitational Wave Observatory (LIGO) spokesperson Gabriela Gon\'{z}ales, ``It takes a village to discover gravitational waves''. This discovery, which entails measurement of displacements over distances as ridiculously small as a thousand times smaller than the diameter of the proton, is undoubtedly the epitome of human knowledge and endeavour.
In the year 1915 Einstein came up with the theory of general relativity, which is the best classical theory of gravity available to date. Einstein's theory has passed every experimental or observational test it has ever been subjected to.
One of the predictions of general relativity was gravitational waves. In fact, Einstein himself came up with this prediction merely a year later in 1916. According to general relativity, gravity is described by the curvature of spacetime. Matter tells spacetime how to curve and spacetime tells test bodies how to move. Spacetime is dynamical unlike in the Newtonian view of the world. Gravitational waves are the tiny ripples on the fabric of spacetime which travel at the speed of light generated by the accelerating masses. When the spacetime metric $g_{\mu\nu}$ is written as $g_{\mu\nu}=\eta_{\mu \nu} + h_{\mu \nu}$, where $\eta_{\mu \nu}$ is the flat metric expressed in Cartesian coordinates and $h_{\mu \nu}$ is the tiny perturbation around it and the vacuum Einstein equations are linearised in $h_{\mu \nu}$, we get the wave equation $\Box h_{\mu \nu}=0$.
to the leading order, in the so-called TT gauge. Here $\Box$ is the flat space d\'{}alembertian. Thus, it implies that the metric perturbation $h_{\mu \nu}$ represents the wave that travels at the speed of light and it is known as the gravitational wave \cite{carroll,schutz}. With the recent discovery of gravitational waves, Einstein's theory of general relativity has survived yet another round of intense scrutiny and passed an~experimental test with the flying colours.
Like electromagnetic waves, gravitational waves also admit two polarizations and are transverse to the direction of propagation. Consider a gravitational wave travelling along \emph{z}-axis in the Cartesian coordinates in flat spacetime. If it is linearly polarized along \emph{x}-axis, the polarization state is referred to as plus polarization. If it is linearly polarized along the direction oriented at an~angle of 45$^{\circ}$
to the \emph{x}-axis in \emph{x}-\emph{y} plane, the polarization state is referred to as cross-polarization. The effect of the gravitational wave with plus polarization increases the proper lengths along \emph{x}-axis and decreases the proper lengths along \emph{y}-axis during the first half-cycle i.e., during the crest, whereas it decreases the proper lengths along the \emph{x}-axis and increases the proper lengths along \emph{y}-axis during the second half-cycle i.e., during the trough. For the cross polarization the same pattern is repeated but is rotated by 45$^{\circ}$ in the \emph{x}-\emph{y} plane (see Figure \ref{fig1}). The strength of the gravitational wave, for instance with plus polarization, denoted by $h$, is a dimensionless number referred to as a gravitational wave strain and is defined as the difference of the fractional change in proper lengths along \emph{x}-axis and \emph{y}-axis, i.e., $h= \left( \frac{\delta L_{x}}{L_{x}}- \frac{\delta L_{y}}{L_{y}} \right)$. Gravitational wave strain is generally extremely small owing
to the extreme weakness of gravity as compared to other forces. For instance, the peak gravitational wave strain during the first detection event was as small as $10^{-21}$. Thus, it is difficult to detect the gravitational waves.
In fact, Einstein himself thought that the gravitational waves would never be detected. It took elaborate efforts over many decades to devise an experimental setup with immense sensitivity, namely the few-km-long Michaelson interferometer, that could directly detect the gravitational waves.

The process of generation of gravitational waves by the non-relativistic source can be understood by solving the linearized Einstein equation with source written in the Lorenz gauge, namely $\Box \bar{h}_{\mu \nu}=-\frac{16 \pi G}{c^4}T_{\mu \nu}$,  where $\Box$ is again the d\'{}alembertian, $\bar{h}_{\mu\nu}$ is the trace reversed part of the metric perturbation, and $ T_{\mu\nu}$ is the stress-tensor associated with the non-relativistic source. The solution of this equation is given by $\bar{h}_{ij}= \frac{2G}{c^4 r}\ddot{I}_{ij}(t_{r})$ where $r$ is
the distance from the source and $I_{\mu\nu}$ is the mass quadrupole moment of the source computed at the retarded time $t_{r}=t-\frac{r}{c}$. This formula implies that the gravitational wave produced by the non-relativistic source is proportional to the second time derivative of its mass quadrupole moment. The rate of loss of energy $P$ of the source via emission of gravitational radiation depends on the square of third time derivative of the quadrupole moment and is given by $P=-\frac{G}{5c^4} \langle \dddot{I}_{ij}\dddot{I}^{ij} \rangle$, where the bracket represents the time average. The validity of this formula was verified by the observation of the Hulse–Taylor binary pulsar,
which consists of two neutron starts going around each other in an orbit, losing orbital energy slowly due to the emission of gravitational waves leading to the shrinking of the orbit and increase in the orbital period which can be inferred from the variation of time of arrivals of the radio pulses. This binary system provided indirect evidence of the existence of the gravitational waves and validated the general relativity. To go beyond the non-relativistic regime
and linear approximation where the source exhibits strong gravitational field and motion of source is relativistic or where the higher order corrections are relevant for the description of the physical system, one has to resort to elaborate techniques such as black hole perturbation theory, numerical relativity and post-Newtonian theory.
In fact, one of the main achievements of the discovery of gravitational waves
is the possibility to compare the numerically calculated waveforms in the full nonlinear regime with the actual observations. As a matter of fact, Einstein himself denied the existence of gravitational waves in full non-linear theory. The first proper theoretical treatment was given by Trautman and Robinson \cite{Trautman,Robinson}.

There are many sources of astrophysical origin that emit gravitational radiation which can potentially 
be detected with the current detector sensitivities. These sources include black hole - black hole binary, black hole – neutron star binary and neutron star – neutron star binary systems, supernova explosions, rapidly rotating deformed neutron stars, inflation, phase transitions in early universe and dynamics of cosmic defects such as cosmic strings \cite{satya}.

The most prominent sources for the ground-based gravitational wave interferometric detectors are the binary systems made up of compact objects such as black holes and neutron stars. In fact, the first detection of gravitational waves is associated with the binary black hole system. Compact objects go around each other in a quasi-circular shrinking orbit for a long time, emitting gravitational waves and eventually merge to form a single black hole. The gravitational wave signal of neutron star binary is slightly different from that of the black hole binary during the orbital evolution due to the fact that the neutron stars are distorted due to the tidal interaction in the later stage when they are sufficiently close. This deformation, which depends on the equation of state of matter constituting the neutron star, is imprinted on the gravitational wave signal. The waveform during the merger phase is
also quite different since the merger of neutron star gives rise to the hypermassive neutron star which radiates for some time before it collapses to form a black hole.

Another source of the gravitational radiation is supernovae. Massive stars, at the end of their life cycle when they run out of nuclear fuel ceasing the process of nuclear fusion and gravity takes over, die catastrophically. Their core undergoes a gravitational collapse, leading to the formation of a neutron star or a black hole, while the outer layers are blown apart constituting a supernova. If~the collapse of the core is asymmetric it leads to the production of gravitational waves in the form of a~burst. The~waveform of the burst radiation is not very well known
at present. If the supernovae take place in our galaxy or galaxies nearby we should be able to detect the burst signal.

Deformed rotating neutron stars with the mountain on their crust as high as a few cm emit continuous gravitational waves. The signal is almost a pure sinusoid at twice the frequency of the rotation of a neutron star in the source frame with a tiny negative frequency derivative. Although the signal is expected to be weaker than that due to compact binary coalescence, it lasts for a very long time. A continuous signal will allow us to constrain various properties of neutron stars such as nuclear matter equation of state. Although so far there is no detection, the spin-down limit has been beaten for Crab and Vela pulsars limiting the loss of rotational energy of pulsars due to the emission of gravitational waves to a small fraction .

Stochastic gravitational radiation is emitted by various physical processes that are inherently stochastic in nature
or due to the incoherent admixture of various coherent signals that are weak enough and are not resolved individually.
For instance, the inflation which is the phase of accelerated expansion in early universe at nearly constant Hubble rate gives rise to a stochastic gravitational wave background. The origin of gravitational radiation during the inflation is inherently quantum mechanical. Apart from inflation, various phase transitions in the early universe, oscillatory dynamics of cosmic defects such as cosmic strings are also expected to rise to such a signal. Stochastic
gravitational wave background due to the unresolved binary black hole mergers could be detected in the upcoming years with ground based detectors.

Gravitational waves are also expected to surprise us, providing access to completely new objects and phenomena.

\section{Gravitational Wave Detectors}

The quest for the detection of gravitational waves started in the 1960s with the pioneer efforts of Weber.
He tried to detect gravitational waves with heavy metal bars that had a size of around 1 m. A gravitational wave passing through the bar would excite it into oscillations due to its tidal effects which could in principle be measured. The bar detectors had low sensitivity essentially due to their small size. Moreover, sensitivity peaked sharply around the resonance making the detectors narrow band. In fact Weber claimed a detection, but his results could not be replicated by other groups \cite{satya}.

\begin{figure}
\begin{center}
\includegraphics[width=12 cm]{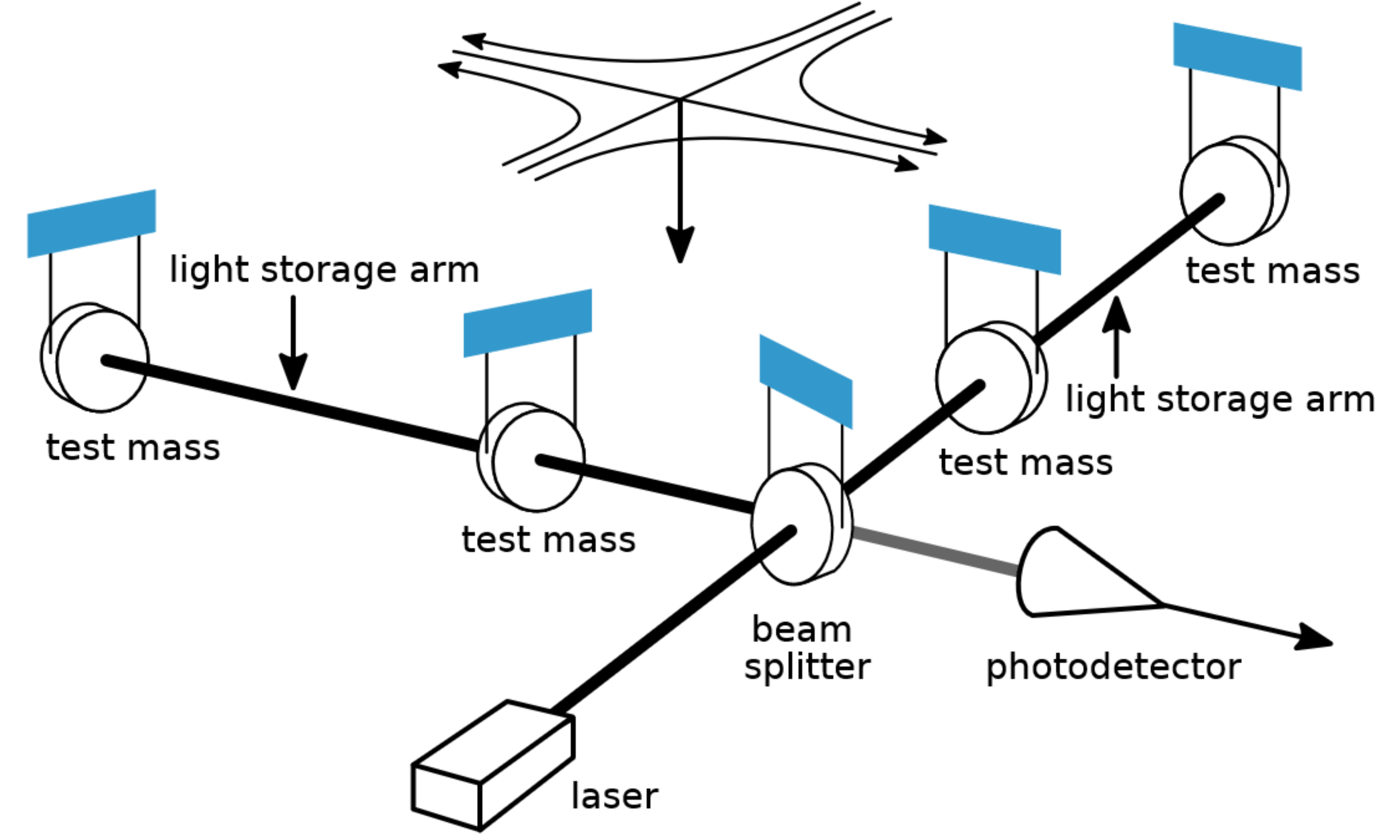}
\caption{ This is a schematic diagram of the Michaelson interferometer. A monochromatic laser light is
split into two beams by the beam splitter which travel along the two perpendicular arms. The laser light moves back and forth in the two arms between the two mirrors depicted as test masses in the figure and is then made to combine
again to form an interference pattern. A gravitational wave passing through the interferometer in the direction perpendicular to its plane for instance, will change the length of one arm with respect to another, causing relative phase shift of the laser light, resulting into the shift in the interference pattern. (Image credit: the Laser Interferometer Gravitational Wave Observatory (LIGO)) }\label{fig2}
\end{center}
\end{figure}

These days we use Michaelson interferometers as gravitational wave detectors (See Figure \ref{fig2}). A monochromatic laser light with wavelength 1064 \emph{A}$^{\circ}$ is used. The laser beam is split into two by the beam splitter. Two beams move back and forth in two arms between the two mirrors suspended vertically at the end of each arm which are free to move back and forth in the horizontal direction. Two laser beams are then combined together to form the interference pattern.
When the gravitational wave passes through the detector it stretches one arm of the interferometer with respect to the other as described earlier. The gravitational strain $h$, the appropriately defined projection of gravitational wave onto the detector, is the relative change of length of the two arms. Phases of the beams passing through the two arms are no longer the same, but suffer from a relative phase shift as a result of which the interference pattern shifts, providing the signature of the gravitational wave signal. Thus, the working principle of the
interferometric gravitational wave detector. For a fixed gravitational wave strain, the~larger the length of the arm,
the larger the change in the length of the arms of the interferometer, making the detection easier. Thus, the effort is made to make the detector arms sufficiently long. Furthermore, as stated earlier, light bounces back and forth
between the mirrors in the detector arm, making the effective length larger and thus enhancing the sensitivity of the detector.

\begin{figure}
\begin{center}
\includegraphics[width=17 cm]{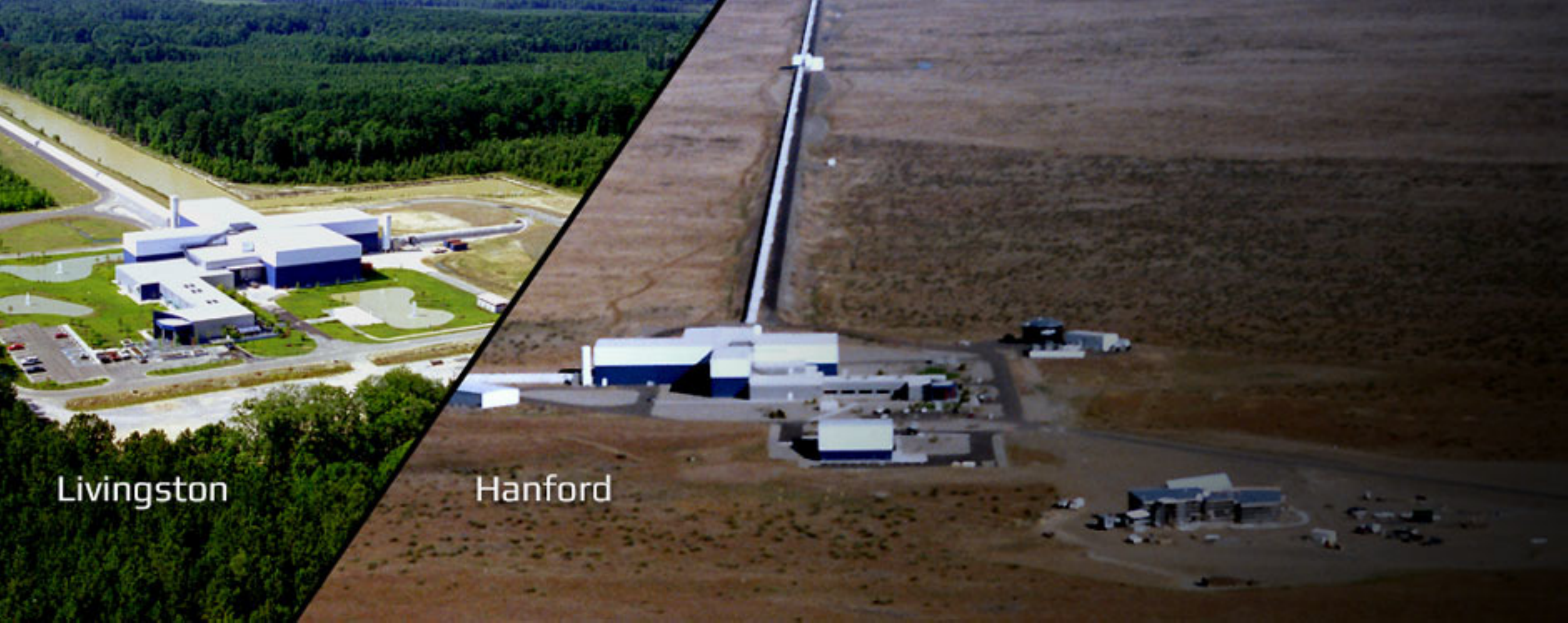}
\caption{ This is the picture of the Laser Interferometer Gravitational Wave Observatory, abbreviated as LIGO.
There are two twin detectors in Livingston, Lousiana and Hanford, Washington separated by 3000 km. The arm length is 4 km. (Image credit: LIGO) }\label{fig3}
\end{center}
\end{figure}

There are two laser interferometers currently in operation. They form part of the Laser Interferometer Gravitational Wave Observatory, abbreviated as LIGO. One is located in Livingston, Lousiana and other is located in Hanford, Washington (See Figure \ref{fig3}). Their armlength is 4 km and they are located 3000~km apart. They were operational in their initial configuration between 2002 and 2010. There were no detections during this period, but non-detection allowed us to put interesting astrophysical bounds on various parameters such as merger rates for black hole binaries. LIGO started functioning in the advanced configuration in 2015 and the first detection miraculously happened almost immediately after the detectors were turned on. The detectors will be upgraded and will reach their design sensitivity in 2019. There is a 3-km detector in Europe near Pisa, Italy called Virgo which will be operational in its advanced configuration later this year. There is also a 600-m-long detector called GEO near
 Hannover in Germany which is operating at this moment. A Japanese underground detector known as KAGRA which is 3 km in armlength is being commissioned right now and would presumably start operating from the year 2019. A third LIGO detector is planned to be built in India which will potentially start functioning in 2024. Unlike the bar detectors, interferometric detectors are broadband detectors. LIGO detectors are sensitive to the gravitational wave signal in the frequency range from 10 Hz to a few kHz that have amplitude of the order of $h\approx 10^{-21}$ which corresponds to the displacement of mirror of the order of $10^{-18}$ m, almost 3 orders of magnitude smaller that the diameter of the proton. The next-generation detectors such as the Einstein Telescope to be built in Europe and the Cosmic Explorer to be built in the USA will have a significantly better sensitivity than the current detectors.

The space-based detector Laser Interferometer Space Antenna, abbreviated as LISA, will be launched in the year 2034 by the European Space Agency. It will overcome the barrier faced by the ground-based detectors, namely the seismic noise at lower frequency. It will consist of three satellites in the triangular shape in the heliocentric orbit that would exchange laser light between them. The distance between the satellites i.e., the armlength of the detector, would be of the order of millions of km. LISA would be sensitive to the gravitational radiation in the frequency range $10^{-4}$ Hz to $10^{-1}$ Hz. While the main target of the ground-based detectors would be stellar mass black holes and neutron stars, LISA would aim to detect supermassive black holes and extreme mass ratio inspirals. In 2016, a mission named the LISA-Pathfinder was launched to test the technology associated with LISA and its results surpassed expectations.

A gravitational wave passing through the region between the earth and the distant pulsar would modify the time of arrival of pulses. Thus, the quadrupolar correlation described by the Hellings and Downs curve of arrival times for a bunch of the most stable known pulsars would imply the presence of gravitational radiation. This technique of detection of gravitational waves is known as Pulsar Timing Array or PTA. PTA would be sensitive in the frequency range $10^{-9}$ Hz to $10^{-8}$ Hz and would detect the gravitational radiation from the population of inspiralling supermassive binary black holes.

In order to detect the primordial stochastic gravitational waves from the inflation one has to resort to cosmic microwave background or CMB. The primordial gravitational waves would generate B-mode polarisation or the vorticity
in the polarization pattern which could be detected by several of dedicated CMB experiments. However the astrophysical background must be taken into account carefully while carrying out the analysis. In 2014 there was a claimed detection by BICEP2 detector which was refuted later by PLANCK mission.

\section{The First Detection of Gravitational Waves}

\begin{figure}
\begin{center}
\includegraphics[width=10 cm]{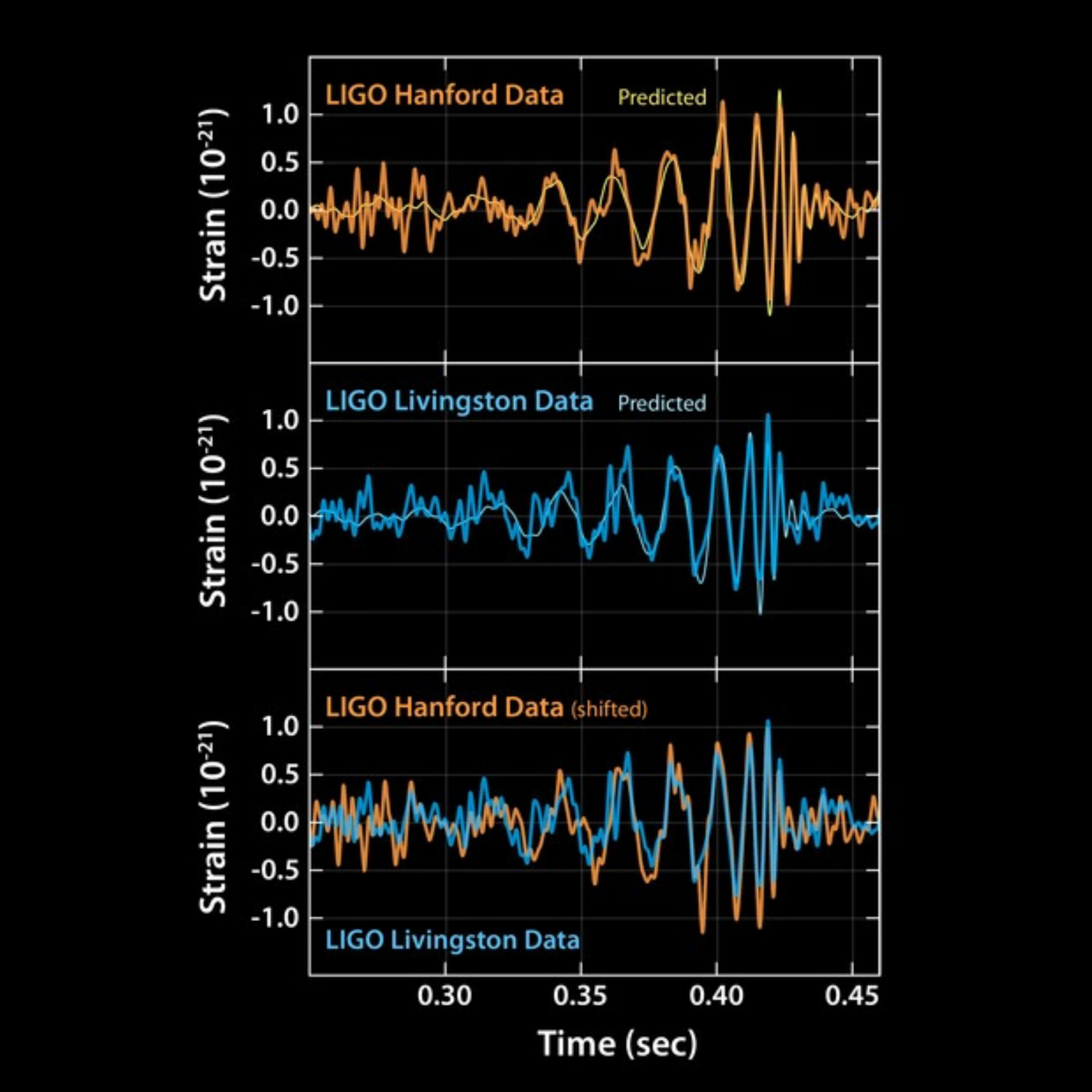}
\caption{ The gravitational waveforms in this figure correspond to the first detection event
GW150914. The gravitational wave strain is plotted against time. The top two panels show the detector
output from the LIGO Hanford and LIGO Livingston. The waveforms predicted from general relativity are
superimposed on the detector output. The bottom panel shows the detector output
from the LIGO Livingston detector along with that from LIGO Hanford detector shifted appropriately
taking into account the time lag in the arrival of signal and the difference in the detector orientations.
They match well indicating that the same signal was detected in both the detectors. (Image credit: LIGO) }\label{fig4}
\end{center}
\end{figure}

The advanced LIGO detectors at the Livingston and Hanford sites were turned on in September 2015 and the
first detection occurred almost immediately thereafter on 14 September 2015, around 09:50 UTC.
 The gravitational wave passed through the Livingston detector initially and then through the Hanford detector 7 ms later accounting for the light travel time between the two spatially separated locations. The signal was identified by the online search
pipeline merely 3 min after the event. The~offline analysis was carried out later confirming the detection and the parameter estimation was performed. The signal to noise ratio was in fact as high as 24.
The peak amplitude of the gravitational wave strain was $h=10^{-21}$ as can be inferred from Figure \ref{fig4}. The band-passed and notched filtered data output from both the detectors is shown in the top two panels of Figure \ref{fig4}. The detector output from the two detectors is superimposed after appropriate inversion taking into account the light travel time between the two detector locations and the different orientation of the arms. It matches very well, indicating the presence of the same signal in both the detectors. The signal was generated by the merger of two stellar mass black holes. Masses of the individual black holes were estimated to be 36 $ M_{\odot}$ and 29~$ M_{\odot}$. In contrast, the mass of the remnant black hole formed as a result of merger was 62 $ M_{\odot}$ and it has a~dimensionless spin of 0.7. The amount of energy emitted in the gravitational radiation was worth 3~$ M_{\odot}$. The distance to the source was 1.3 billion light years \cite{detection1}. The peak luminosity of the event was $3 \times 10^{56}$ erg/sec, making it the most luminous event exceeding the integrated luminosity of all the stars taken together in the observable universe.

Despite the effort to minimize the noise and enhance the sensitivity of the detector in the best possible way, the noise generally exceeds the signal in the detector output. The gravitational signal that is hidden in the noise must be excavated out using optimal data analysis techniques such as matched filtering where the expected signal is correlated with the detector output. Templates which represent all possible gravitational wave signals are used for this purpose. Thus, it is essential to have an accurate waveform a priori. In the case of binary black hole coalescence which is the most promising source of gravitational radiation for ground-based detectors and also the source of the signal detected by LIGO, it is essential to solve two-body problem in general relativity. It is a difficult problem unlike the Newtonian case where the exact solution is known. Various sophisticated techniques have been developed over any decades to deal with the two-body problem in general relativity.

Initially, two black holes orbiting one another in a quasi-circular orbit are far apart, the gravity is weak and the speeds are non-relativistic. The orbit slowly shrinks due to the emission of gravitational waves which steals the orbital energy. This phase is known as inspiral. During the inspiral, since there is a small dimensionless parameter
in the problem, namely the ratio of orbital velocity $v$ and speed of light $\frac{v}{c}$ , one can use the perturbation theory. The Einstein equations are expanded systematically in powers of this dimensionless parameter and solved order by order. This method is known as post-Newtonian expansion. So far post-Newtonian corrections up to
4 PN order i.e., up to the order $\left(\frac{v}{c}\right)^8$ have been computed.
 Post-Newtonian corrections include various interesting effects such as spin-orbit coupling i.e., coupling between the spins of the black holes and the orbital angular momentum, and tail effects which account for the back-scattering of the gravitational radiation due to the spacetime curvature etc.

When the black holes come sufficiently close, the velocities are relativistic and gravity is strong, and post-Newtonian approximations becomes less reliable. Eventually, black holes plunge towards each other and collide at the velocity close to the speed of light. The black holes merge together to form a~single remnant black hole. This phase is known as merger. A burst of radiation is given out during the merger with often the luminosity of the gravitational waves exceeding the integrated luminosity of all stars in the entire observable universe. As stated earlier, the peak luminosity in case of GW150914 was $3 \times 10^{56}$ erg/sec. Analytical perturbative and approximation techniques fail during the merger and one has to deal with the full non-linear dynamics of general relativity in its full glory using numerical relativity. Full-fledged numerical relativity simulations of the binary black hole merger were carried by many groups out after the initial breakthrough in the year 2005 in this field \cite{Pretorious}.

\begin{figure}
\begin{center}
\includegraphics[width=8 cm]{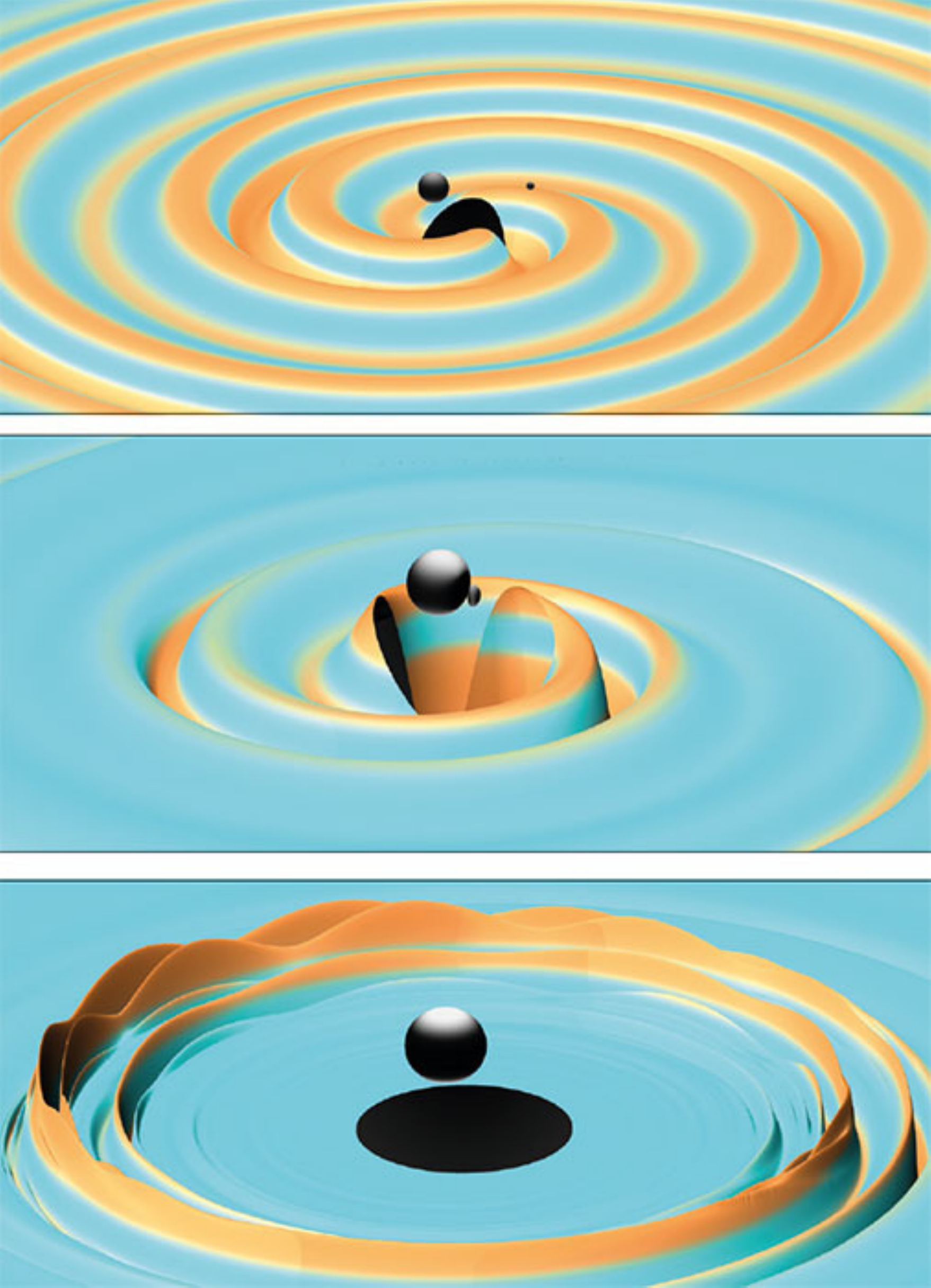}
\caption{ The three stages of binary black hole coalescence, namely the inspiral, merger and ringdown are depicted
in this figure from top to bottom in the context of numerical relativity simulation of GW150914 carried out at Albert Einstein Institute, Germany. In the inspiral phase depicted in the top panel two black holes orbit around one another in a quasi-circular orbit which decays as the gravitational waves slowly drain the orbital energy. During the merger phase depicted in the middle panel two black holes bang into each other at the relativistic speeds and form a single black hole. A~burst of gravitational waves with extremely large luminosity is given out during the merger phase. During the ringdown phase depicted in the lower panel, the remnant black hole which is in the excited phase sheds out the deformation away from the Kerr configuration by emitting gravitational waves. (Image credit: AEI) }\label{fig5}
\end{center}
\end{figure}

The final black hole that is formed as a result of merger is often in the excited state. The~deformations of the black hole away from the stationary Kerr configurations are radiated away. The perturbations die down due to the emission of gravitational waves and the waveform looks like the superposition of damped sinusoids. This phase is known as the ringdown. Techniques of black hole perturbation theory are used to deal with the ringdown, where the metric is written as the sum of Kerr metric and perturbations and Einstein equations are expanded in terms of perturbations and solved order by order. Gravitational waves are emitted at specific discrete values of complex frequencies known as quasi-normal modes. Frequencies of quasi-normal modes depend on the mass and spin of the final Kerr black hole. The GW150914 signal was consistent with the presence of least damped $l=2,m=2$ quasi-normal mode for the estimated mass and spin assuming Kerr hypothesis and general relativity. The snapshots of inspiral, merger and ringdown phases of binary black hole coalescence in the numerical relativity simulation of GW150914 are depicted in Figure \ref{fig5}.

The techniques of effective one body are also used to generate the waveform \cite{Damour}. The~post-Newtonian theory is extended into the strong field regime using the resummation techniques and it is complemented with the input from the numerical relativity simulations. It provides us a full waveforms for the binary black hole coalescence in the analytical form that are accurate and efficient. The detector output for GW150914 in the LIGO Livingston and Hanford detectors superimposed with the waveform of binary black hole coalescence predicted from general relativity using the techniques mentioned above are shown in Figure \ref{fig4}.

There was a second detection of gravitational waves from the binary black hole merger which happened on 26 December
2015. This event is referred to as GW151226. Two black holes with the masses 14 $M_{\odot}$ and 8 $M_{\odot}$ merged to form a remnant black hole of the mass 21 $M_{\odot}$. The signal-to- noise ratio was 13 \cite{detection2}. There was yet another candidate with low statistical significance and hence it is not claimed to be the detection.

\section{Electromagnetic Follow-Up}

\begin{figure}
\begin{center}
\includegraphics[width=9 cm]{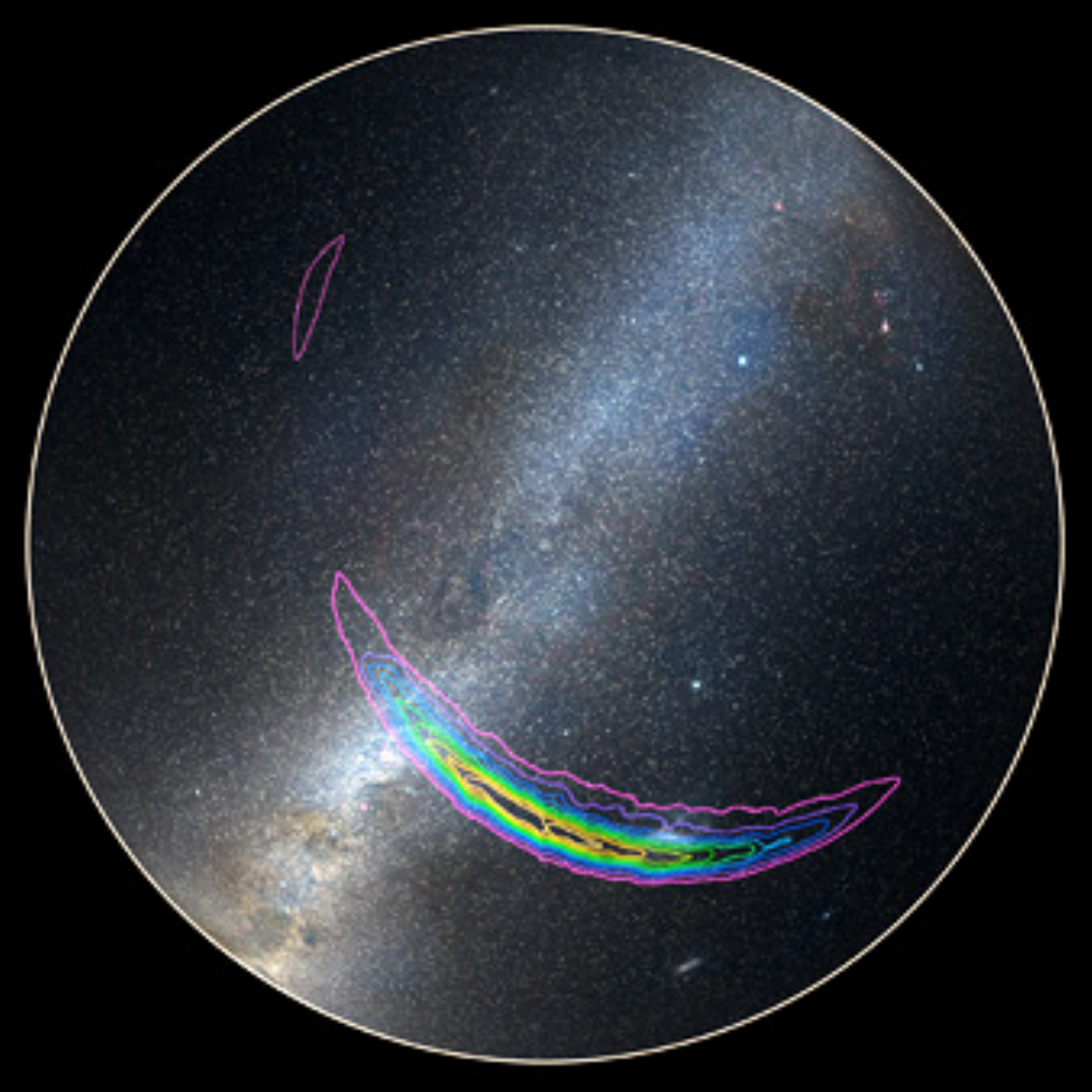}
\caption{ The sky localisation of GW150914 is depicted in this figure. The source of gravitational radiation could be spread over several 100 sq.deg. This is a large region in the sky compared to the area spanned by sun which is merely 0.5 sq.deg.  (Image credit: CalTech) }\label{fig6}
\end{center}
\end{figure}

The sky localization of gravitational wave event GW150914 was carried out from the information of
the time delay of arrival of the signal between the two LIGO detectors and their directional sensitivities.
The sky localisation of this event is depicted in Figure \ref{fig6}. The source could be located over a region in sky that is spread across the area as large as few 100 square degrees
This is pretty large as compared to for instance the sky spanned by sun which is merely 0.5 100 square degrees.
This makes the electromagnetic follow-up rather challenging. A lot of effort was put into search for an electromagnetic counterpart to the gravitational wave event by multiple ground based and space based telescopes all across the electromagnetic spectrum ranging from radio to gamma rays. However, no promising candidate was discovered \cite{em}. This is in line with the expectation that the black hole coalescence events are not expected to harbour an electromagnetic counterpart.

The gravitational wave signal from the binary neutron star merger is expected to be observed in the future. The event of binary neutron star merger could be interesting from the point of view of joint electromagnetic-gravitational wave observation. A binary neutron star merger is believed to be the progenitor of a short gamma-ray burst, which is an energetic event. With the enhancement of sensitivity of the gravitational wave detectors and with multiple detectors going online in the near future, the~sky localization of the gravitational wave event would improve significantly. This~would make the electromagnetic follow-up significantly easier. Observation of short gamma-ray burst coincident with the gravitational wave detection of neutron star merger will allow us to confirm the alleged
link between the two.

\section{Tests of General Relativity}

All the tests of general relativity carried out so far which include solar system tests and tests concerning binary pulsars, deal with the regime where the gravitational fields are weak, velocities are small, and dynamics are quasi-static. For the first time, gravitational wave observation of binary black hole coalescence provided us with an opportunity to probe gravity and test general relativity in the strong field, large velocity and highly dynamical regime which was inaccessible before. Various tests to check the validity of general relativity were carried out as we describe below \cite{gr}.

The best-fit general relativity waveform was subtracted from the detector output and it was checked whether the residual was consistent with pure noise or there was leftover power. It was found that the residual was indeed consistent with Gaussian noise. General relativity was tested to $4 \%$ level using this test, in other words the correlation between the detector output and the waveform based on general relativity is greater than $96 \%$.

\begin{figure}
\begin{center}
\includegraphics[width=12 cm]{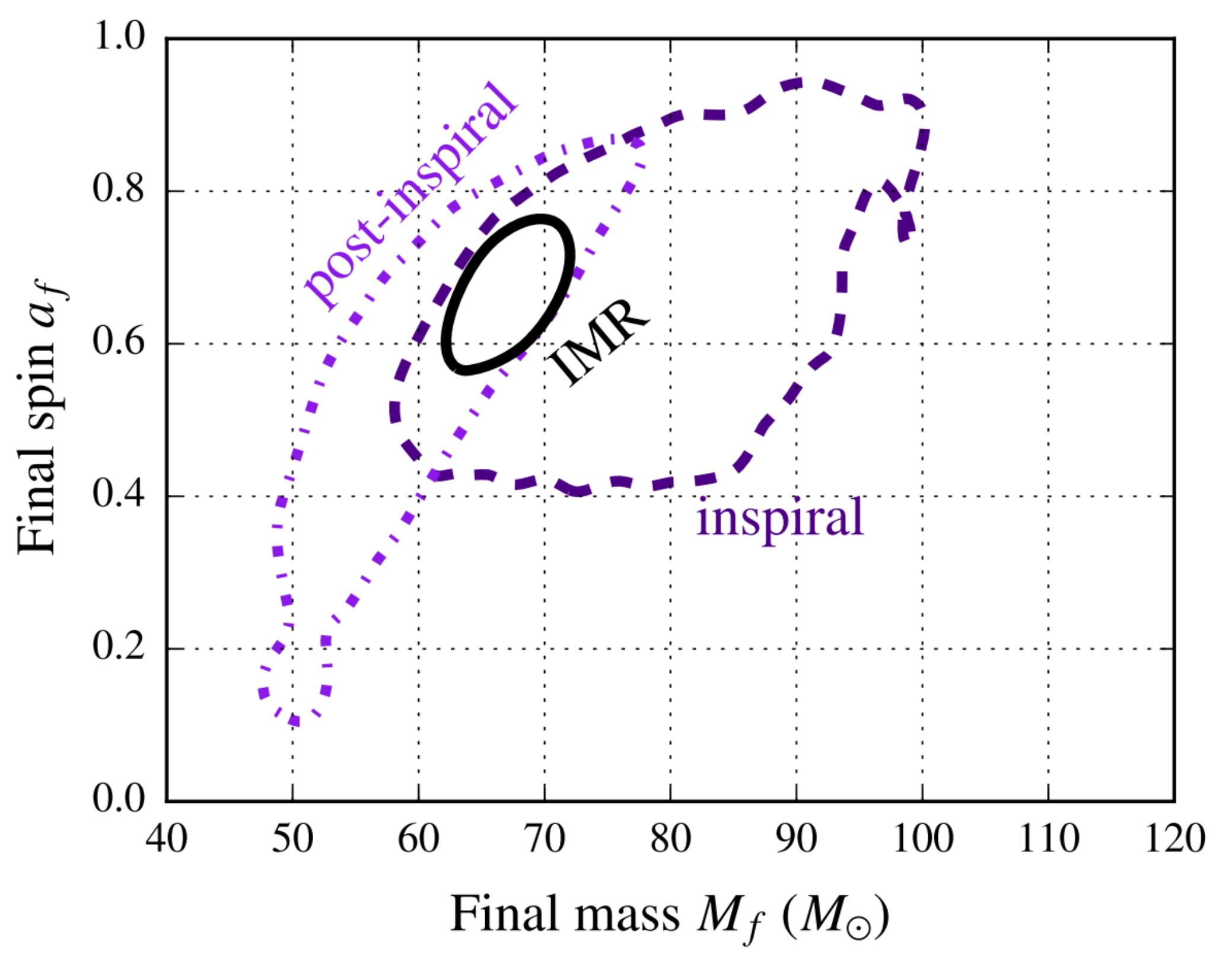}
\caption{ The final dimensionless spin $a_{f}$ is plotted against the final mass $M_f$ in units of solar mass. The dark purple curve confines the $90 \%$ confidence region of the prediction of the final mass and spin based on the general relativistic numerical relativity simulation starting from the information from the inspiral part of the gravitational wave signal. The dotted purple curve encloses $90 \%$ confidence region of the prediction of final mass and spin from the post-inspiral part of the gravitational wave signal. The solid black curve represents the prediction of the final mass and spin based on the full inspiral-merger-ringdown waveform. There is a significant overlap.
(Image credit: LIGO) }\label{fig7}
\end{center}
\end{figure}

The final mass and final dimensionless spin parameters of the remnant black hole, $M_{f}$ and $a_{f}$, can be determined in two ways, by using the inspiral or low frequency part and by using the post-inspiral or high frequency part of the signal. In the case of GW150914 the critical frequency was $132$ Hz. In~order to determine the mass and spin of the remnant black hole input from the numerical relativity is required. Numerical relativity evolution starting from the information of the inspiral phase allows us to predict the final mass and spin of the remnant black hole assuming general relativity. The dark purple contour in Figure \ref{fig7} confines the $90 \%$ confidence region in the $M_{f}-a_{f}$ plane based on the prediction using inspiral part of gravitational waveform and numerical relativity. On the other hand, the dotted purple curve confines $90 \%$ confidence region based on the post-inspiral part of signal. The post-inspiral part of the signal contains merger and ringdown. Again using the numerical relativity fitting formulae
the confidence region is inferred. It is clear from Figure \ref{fig7} that two contours show significant overlap confirming that the two independent predictions of general relativity are indeed mutually consistent.

\begin{figure}
\begin{center}
\includegraphics[width=12 cm]{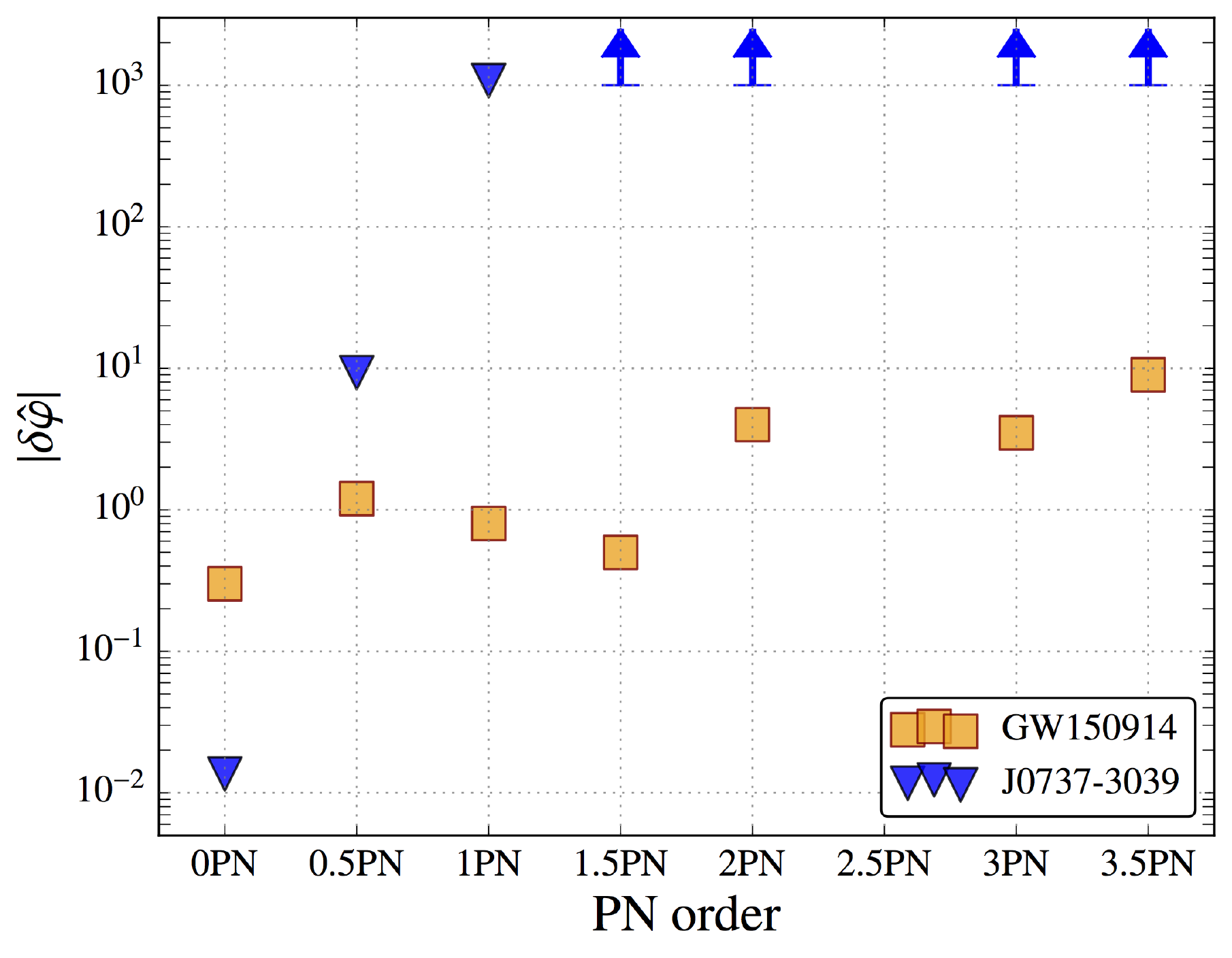}
\caption{ Upper bounds (denoted by $|\delta \hat{\phi}|$) on the fractional deviations of the post-Newtonian coefficients from their general relativistic prediction are plotted against the PN orders for the gravitational wave observation as well for binary pulsar observation. Binary pulsar puts better constraints at the leading order,
while bounds are much better at the higher orders for gravitational wave observation. (Image credit: LIGO) }\label{fig8}
\end{center}
\end{figure}

As described before, post-Newtonian approximation, where we expand the Einstein equation perturbatively in powers of orbital velocity, allows us to compute the inspiral part of the gravitational wave signal to the desired accuracy.
The waveform can be presented in the Fourier domain and the phase is represented as post-Newtonian series. The coefficients at each order are fixed by general relativity. We modify those coefficients
to mimic possible deviations from general relativity.
 Figure~\ref{fig8} shows upper bounds on the fractional deviations from the general relativistic predictions $|\delta \hat{\phi}|$, from~gravitational wave and binary pulsar observations. The upper bound is quite strong at the leading order from the binary pulsar observations, but bounds are pretty weak at higher orders beyond 1 PN since the orbital velocity is small. Upper bounds imposed by gravitational waves observation at higher orders are significantly better since in the late inspiral regime we have access to higher~velocities.

\begin{figure}
\begin{center}
\includegraphics[width=12 cm]{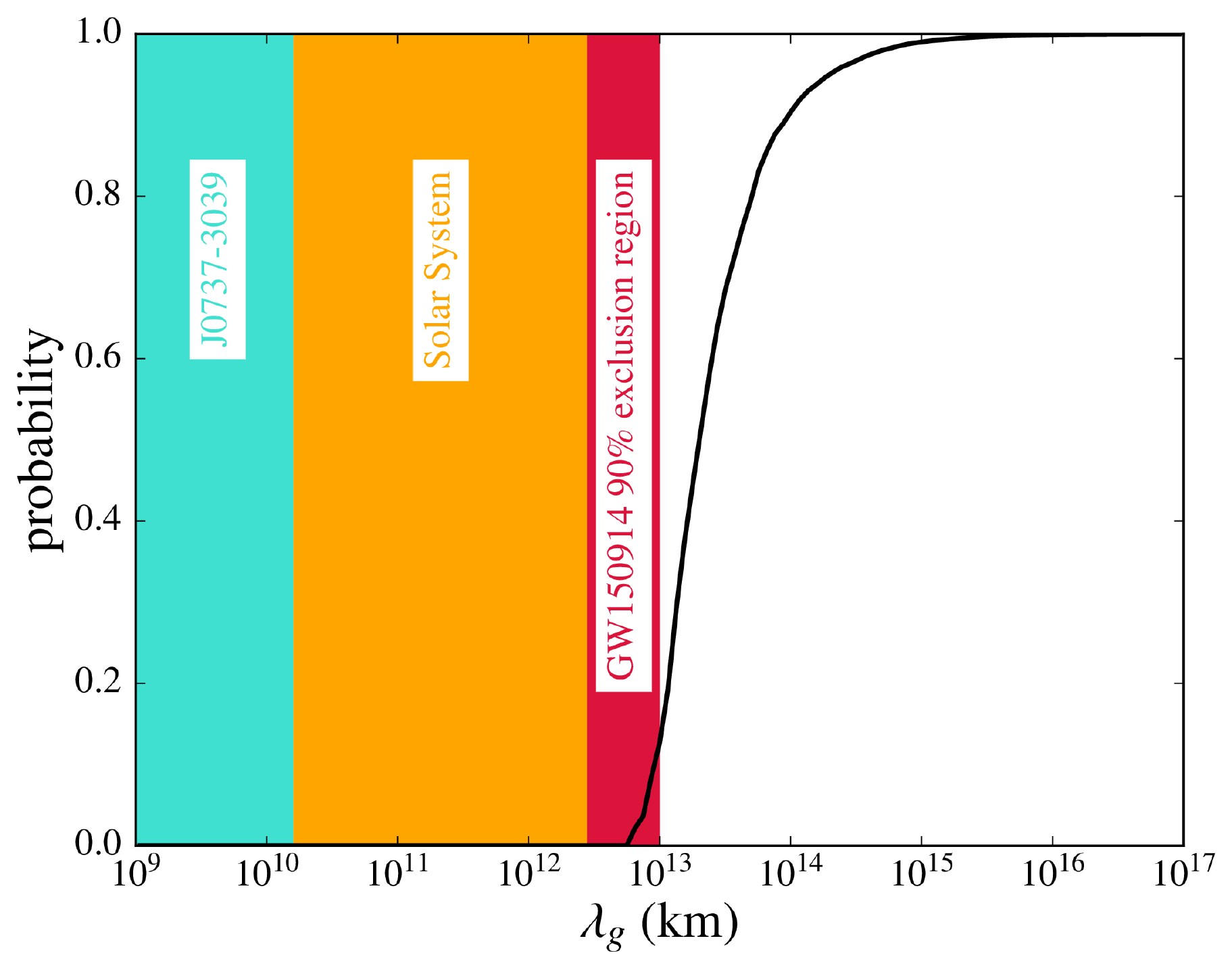}
\caption{ Probability distribution function for the Compton wavelength of the graviton $\lambda_{g}$ expressed in km is plotted in this figure. The gravitational wave puts better constraints than solar system tests of gravity. (Image credit: LIGO) }\label{fig9}
\end{center}
\end{figure}

Gravitons are massless particles in general relativity and thus the gravitational waves travel at speed of light. At present we do not have access to sensible well-defined theory of massive gravity. However,~one could deal with a hypothetical situation in which gravitons have non-zero fixed Compton length i.e., non-zero mass. Gravitational waves in that case would travel slower than speed of light, moreover, the speed will be different at different frequencies exhibiting non-trivial dispersion relation. The phase of the gravitational wave signal in such a case would deviate from the general relativistic prediction. This allows us to put constraints on the Compton wavelength.
Newton's law of gravity in such a case will also get modified, which allows to put constraints from solar system tests of gravity. Figure \ref{fig9} shows the probability distribution function for the Compton wavelength of graviton. The lower bound on the Compton wavelength of graviton imposed by gravitational wave observation is a factor of 3~better than the one from solar system tests.

All studies carried out to detect deviation from general relativity indicate that GW150914, the~process of binary black hole coalescence, was in accordance with the vacuum Einstein equation of general relativity and no statistically significant deviation was found within the limits such as sensitivity of the detectors and nature of event itself. In the future it would be possible to use better constraints with detectors with better sensitivities due to the availability of the network of detectors, with detections that exhibit higher signal to noise ratio and by combining multiple observations. We should be able to test the no hair theorem, the second law of black hole mechanics i.e., the area theorem, look for extra polarization modes, and so on. The golden era in the field of gravity, fundamental physics and astronomy has just begun with the first detection of gravitational waves (GW150914) and many more exciting developments and discoveries are ahead of us.

\section*{Acknowledgments}
We acknowledge support from the NCN grant Harmonia 6 (UMO-2014/14/M/ST9/00707).

\section*{Author Contributions}
A.K. conceived the paper and provided initial input whereas M.P. wrote the paper.

\end{document}